\documentclass[prd,twocolumn,floatfix,amsmath,nofootinbib,amssymb,floatfix]{revtex4}
\usepackage{graphicx,color,dcolumn,booktabs,bm,multirow}
\usepackage{longtable,lscape}
\usepackage{txfonts}
\usepackage{overpic}
\usepackage{amssymb}
\usepackage{array}
\usepackage{indentfirst}
\usepackage{feynmf}   
\usepackage{bbding}
\usepackage{slashed}  
\usepackage{cases}
\usepackage{color}
\usepackage{multirow}
\usepackage{epstopdf}
\usepackage{tabularx}
\usepackage{graphicx,color,dcolumn,booktabs,bm}
\usepackage[colorlinks,
citecolor=blue,
anchorcolor=red,
menucolor=red,
linkcolor=red,
filecolor=red,
runcolor=red,
urlcolor=blue,
frenchlinks=red]{hyperref}
\usepackage{float}

\begin{document}
\title{Production of $\Xi(1530)$ in the $K^- p$ scattering process}
\author{Quan-Yun Guo$^{1}$}\email{guoquanyun@seu.edu.cn}
\author{Jing Liu$^{2}$}\email{liujing@hue.edu.cn}
\author{Peiwen Wu$^{1}$}\email{pwwu@seu.edu.cn}
\author{Dian-Yong Chen$^{1,3}$}\email{chendy@seu.edu.cn}
\affiliation{$^1$ School of Physics, Southeast University, Nanjing 210094, People's Republic of China}
\affiliation{$^2$ School of Physics and Mechanical Electrical and Engineering, Hubei University of Education, Wuhan 430205, China}
\affiliation{$^3$ Lanzhou Center for Theoretical Physics, Lanzhou University, Lanzhou 730000, China}
	\date{\today}

\begin{abstract}
In the present work, we examine the production of $\Xi(1530)$ in the $K^- p \to K^{+} \Xi(1530)^{-}$ and $K^- p \to K^{0} \Xi(1530)^{0}$ reactions utilizing an effective Lagrangian approach. To accurately fit the cross sections for both processes, we include nine $\Lambda$ and $\Sigma$ hyperons and their resonances in both $s$- and $u$-channel processes. Considering the discrepancy of the measured cross sections for $K^- p \to K^+ \Xi(1530)^-$ within the range $\sqrt{s}=[2.087, 2.168]\ \mathrm{GeV}$, we employ two distinct fitting strategies: a uniform weighting scheme (model A) and a different weighting approach (model B). A comparative analysis suggests that model A yields a superior global agreement with experimental data compared to model B. Beyond fitting the cross sections, we also estimate the individual contributions from various intermediate states. Our results reveal that the cross section arising from the $\Sigma(1193)$ intermediate process is dominant. Furthermore, we predict different cross sections for $K^- p\to K^+\Xi(1530)^-$ and $K^- p\to K^0\Xi(1530)^0$ at several representative center-of-mass energies, providing testable predictions for forthcoming J-PARC experiments.
\end{abstract}
	
\maketitle

\section{Introduction}
\label{sec:Introduction}
$\Xi$ hyperons, composed of one up or down quark and two strange quarks, are particularly interesting members of the baryon family because they provide insights into the dynamics of the strong force in environments with high strange quark content. Due to this characteristic, the production of $\Xi$ hyperons is challenging, and experimental observations are limited compared to $\Lambda$ and $\Sigma$ hyperons, which contain only one strange quark. A review by the Particle Data Group (PDG) reveals that only the ground $S$-wave states, $\Xi(1315)^{0,-}$ with $J^{P}=1/2^+$ and $\Xi(1530)$ with $J^P=3/2^+$, are well established with four-star ratings~\cite{ParticleDataGroup:2024cfk}. Additionally, there are four $\Xi$ resonances with three-star ratings: $\Xi(1690)$, $\Xi(1820)$, $\Xi(1950)$, and $\Xi(2030)$. Of these, the $J^{P}$ quantum numbers of $\Xi(1820)$ have been determined to be $3/2^-$, and the spin of $\Xi(2030)$ has been measured to be $5/2$. The remaining five $\Xi$ resonances, $\Xi(1620)$, $\Xi(2120)$, $\Xi(2250)$, $\Xi(2370)$, and $\Xi(2500)$, are only rated as two or one star~\cite{ParticleDataGroup:2024cfk}, indicating that their existence is only considered to be fairly or even poorly established.

Experimental knowledge of the $\Xi$ hyperon spectrum is severely limited. As highlighted by the Particle Data Group review, a significant portion of the current data on $\Xi$ hyperons derives from low-statistics experiments conducted in the 1960s through the 1980s with $K^-$ beams, and in the 1980s and 1990s with hyperon ($\Sigma^-$, $\Xi^-$) beams~\cite{ParticleDataGroup:2024cfk}. Despite these limitations, advancements have emerged from collider experiments. Several excited $\Xi$ hyperons have been identified in the decay products of charmed baryons. Notably, the Belle Collaboration detected the $\Xi(1690)$ hyperon in the $\Sigma^+ K^-$ invariant mass from $\Lambda_c^+ \to \Sigma^+K^- K^+$ decays~\cite{Lyth:2001nq},  subsequently confirmed by BABAR in the $\Xi^- \pi^+$ channel from $\Lambda_c^+ \to \Xi^- \pi^+ K^+$~\cite{BaBar:2008myc}. The Belle Collaboration also made the initial observation of $\Xi(1620)^0$ decaying to $\Xi^- \pi^+$ through $\Xi_c^+ \to \Xi^- \pi^+ \pi^+$ decays~\cite{Belle:2018lws}. Furthermore, the CLAS Collaboration has contributed through the measurement of $\Xi(1321)^{-}$ hyperon photoproduction via the reactions $\gamma p \rightarrow K^{+} K^{+} \Xi^{-}$ and $\gamma p \rightarrow K^{+} K^{+} \pi^{-} \Xi^{0}$~\cite{CLAS:2004gjf, Price:2004hr}, enabling them to obtain a substantial sample of $\Xi(1321)^{-}$. Using a subsequent, high-statistics dataset for the reactions $\gamma p \rightarrow K^{+} K^{+} (X)$ and $\gamma p \rightarrow K^{+} K^{+} \pi^{-} (X)$~\cite{Guo:2007dw}, the CLAS collaboration measured a mass splitting for the ground state $(\Xi^{-},\Xi^{0})$ doublet, which is $(5.4\pm 1.8)$ MeV ~\cite{Guo:2007dw}.

On the theoretical side, comprehensive investigations of the $\Xi$ spectrum have been undertaken using various approaches. For instance, the spectrum and decay properties of $\Xi$ hyperons have been estimated within the constituent quark model, including relativistic~\cite{Capstick:1986ter} and nonrelativistic versions with harmonic confinement potentials~\cite{Chao:1980em,Glozman:1995fu, Xiao:2013xi} or linear plus Coulomb potentials~\cite{Pervin:2007wa}. Furthermore, Ref.~\cite{Bijker:2000gq} explores the spectrum and decay properties of $\Xi$ hyperons using a string like model, where radial excitations are interpreted as rotations and vibrations of the strings. Masses of negative-parity $\Xi$ hyperons have also been analyzed in large-$N_c$ QCD, considering corrections to order $1/N_c$ and first-order SU(3) symmetry breaking~\cite{Schat:2001xr}. Additionally, the mass spectrum and magnetic moments of $\Xi$ baryon resonances have been investigated within the bound-state framework of the Skyrme model~\cite{Oh:2007cr}.

From the perspective of the productions of $\Xi$ states, the two-body process $K^-N \to K^+ \Xi$ is the only feasible binary reaction, while other projectiles such as  photons, pions, and protons, invariably lead to multibody final states involving three or more particles. The experimental investigations of $K^- p$ scattering began nearly 60 years ago. For example, in Ref.~\cite{Berge:1966zz} the productions of $\Xi^-$ and $\Xi^0$ hyperons in $K^-p$ interactions between 1.05 and 1.7 GeV were measured, and the cross sections were reported to be about $150~\mathrm{\mu b}$ for $\Xi^- K^+$ production and $100~\mathrm{\mu b}$ for $\Xi^0 K^0$ and $\Xi \pi K$ productions, in addition to which the production of $\Xi^\ast(1530)$ near threshold was also observed.  Meanwhile, the $\Xi$-hyperon production in the $K^- p$ interaction at $2.24~\mathrm{GeV}$~\cite{London:1966zz} and  3.5 GeV~\cite{Birmingham-Glasgow-LondonIC-Oxford-Rutherford:1966onr} was measured. After that the production cross sections and differential cross sections of $\Xi(1314)$ and $\Xi(1530)$ in the $K^- p$ scattering were further reported at different incident kaon beam momenta until to 1977~\cite{Trippe:1967wat, Trower:1968zz, Burgun:1968ice, Dauber:1969hg, SABRE:1971pzp, deBellefon:1972rjq, Carlson:1973td, Griselin:1975pa, Briefel:1977bp}. It is worth noting that the High-Energy Reactions Analysis Group in CERN~\cite{Flaminio:1979iz} presented a lot of cross-sectional experimental data about $\Xi(1314)^{0,-}$ and $\Xi(1530)^{0,-}$ in the $K p$ scattering processes in 1979, which provides experimental conditions for our study on the production of $\Xi(1530)$.

In early theoretical research, only a few  works made progress on the production of $\Xi$ hyperons due to the limitation of experimental data. In Ref. ~\cite{Agarwal:1971kb}, the authors explained the backward production features of the reaction $K^{-}p \rightarrow \Xi^{-}K^{+}$ by using the two-meson-exchange peripheral model, in which the cross sections and the angular distributions for the reaction $K^{-}p \rightarrow \Xi^{-}K^{+}$ could be well described by using the two-meson-exchange box diagrams. Within the model based on Regge theory, duality and SU(3) symmetry combined with the known data, the authors in Ref.~\cite{Mir:1981qj} investigated the reaction $\bar{K} N \rightarrow \Xi K$ at high energy. 

In 2011, the production of $\Xi(1314)$ in the $K^{-}N$ scattering process was investigated~\cite{Sharov:2011xq}, where the cross sections and differential cross sections for the $K^-N\to K \Xi(1314)$ reaction were fitted within the phenomenological model. By utilizing the effective Lagrangian approach, the $s$- and $u$-channel exchange by $\Lambda$, $\Sigma$, and their resonances were taken into consideration, and they found that the cross sections and differential cross section could be well described. Similarly, in Ref.~\cite{Jackson:2015dva}, the authors also investigated the process $\bar{K} N \rightarrow \Xi K$ in the effective Lagrangian method. It is worth noting that in addition to the contributions of $s$ and $u$ channel, a contact diagram was also included, and their estimations showed that the dominant contribution came from the contact diagram, while the $s$- and $u$-channel contributions were only about $20\%$ of that of the contact term. Moreover, based on a hybrid Regge-plus-resonance model, the authors of Ref.~\cite{Kim:2023jij} also studied the process $\bar{K} N \rightarrow \Xi K$. The estimations showed that the bump structures in the total cross sections could be described by the $s$-channel hyperon resonances, while the dominant contributions come from $\Lambda(2100)$ and $\Sigma(2030)$ resonances.

In addition to the production of $\Xi(1314)$ in the $K^{-}p$ scattering process, we also notice that there have been some data related to cross sections for $K^- p \to K ^+ \Xi(1530)^-$ and $K^- p \to K^0 \Xi(1530)^0$ in Refs~\cite{Berge:1966zz, Dauber:1969hg, Briefel:1977bp, Flaminio:1979iz}. Moreover, the extension project has been proposed in the J-PARC hadron experimental facility, where the high energy kaon beam with high quality can be available. With the kaon beam, the production of $\Xi$ hyperon, including $\Xi(1530)$, will be possible. Thus, we propose to investigate the processes $K^- p \to K^{+} \Xi(1530)^{-}$ and $K^- p \to K^{0} \Xi(1530)^{0}$ in the present work. By fitting the cross sections of two processes, the model parameters could be determined, and then the differential cross sections are predicted, which can further be tested by the experimental measurements at J-PARC in the future.

This work is organized as follows. After the Introduction, the effective Lagrangian approach employed to investigate the processes $K^- p \to K \Xi(1530)$ are presented. In Sec.\ref{sec:MA}, the numerical results and relevant discussions of the cross sections and the differential cross sections for the considered processes are presented. The last section is devoted to a short summary.

\section{$\Xi(1530)$ production in the $K^- p $ scattering process}
\label{sec:MS}
\begin{figure}[t]
	\begin{tabular}{ccc}
		\centering
\includegraphics[width=75mm]{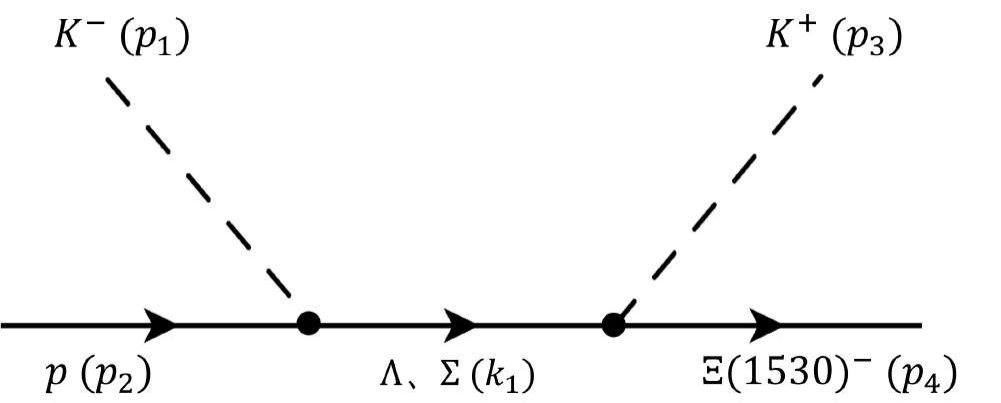}&\\
		{\large(a)} &\\
		\includegraphics[width=75mm]{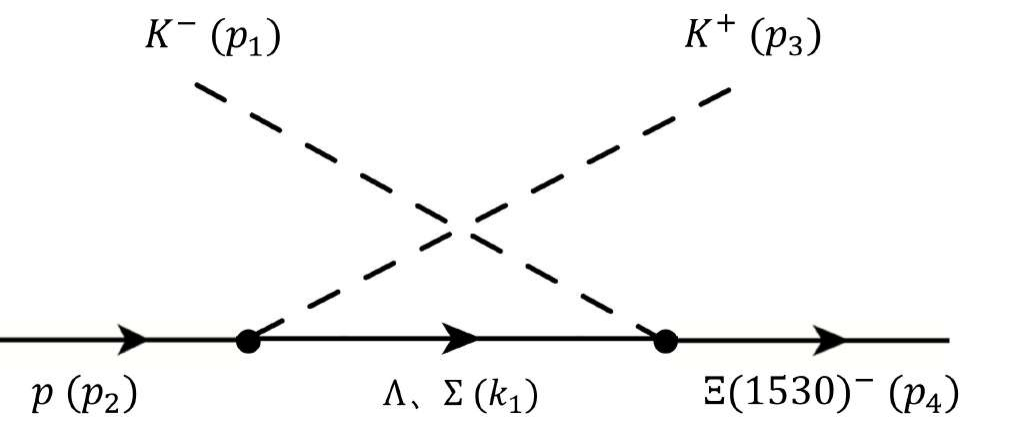}&\\
		\\
		\large {(b)}& \\
	\end{tabular}
	\caption{Diagrams contributing to the process of $K^{-}p \rightarrow K^{+} \Xi(1530)^{-}$, 
		 corresponding to the (a) $s$- and (b) $u$-channel contributions, respectively.}\label{Fig.1}
\end{figure}

In the present work, we employ an effective Lagrangian approach to describe the interactions of hadrons in the $K^- p \to K \Xi(1530)$ process. The diagrams contributing to $K^- p \to K^{+} \Xi(1530)^{-}$ are presented in Fig.~\ref{Fig.1}, where $\Lambda$ and $\Sigma$ hyperons and their resonances could serve as intermediate states. In the following, we assign the notation $Y_{J^P}$ to represent $\Lambda$ and $\Sigma$ hyperons and their resonance states characterized by spin quantum number $J$ and parity $P$. In the present calculations, we only take the lower spin states with $J\leq 5/2$ into consideration, as these states are expected to dominate the physical processes occurring near the production threshold region. The effective Lagrangians for $pY_{1/2^\pm}K$ read~\cite{Jackson:2015dva},
\begin{eqnarray}
\mathcal{L}_{pY_{1/2^\pm}K} &=& (\mp i) f_{pY_{1/2^\pm}K} \bar{p}
\left\{
\begin{array}{c}
	\gamma_5 \\
	1
\end{array}	
\right\} 
Y_{1/2^\pm} K + H.c,
\label{Eq.1}
\end{eqnarray}
where the upper and lower symbols in the curly bracket are the positive and negative parity of the $Y$ hyperon, respectively. 
For $pY_{3/2^\pm}K$ and $pY_{5/2^\pm}K$ couplings, the effective Lagrangians are~\cite{Jackson:2015dva},
\begin{eqnarray}
\mathcal{L}_{pY_{3/2^\pm}K} &=& \frac{f_{pY_{3/2^\pm}K}}{m_\pi} \bar{p} 	
\left\{
\begin{array}{c}
	1 \\
	\gamma_5
\end{array}	
\right\} 
Y^\mu_{3/2^\pm } \partial_\mu K +H.c.\nonumber \\
\mathcal{L}_{pY_{5/2^\pm}K} &=& (-i) \frac{f_{pY_{5/2^\pm}K}}{m^{2}_\pi} \bar{p} 	
\left\{
\begin{array}{c}
	\gamma_5 \\
	1
\end{array}	
\right\} 
Y^{\mu \nu}_{5/2^\pm } \partial_{\mu} \partial_{\nu} K +H.c,
\end{eqnarray}
respectively. In addition, for $\Xi(1530)Y_{J^P}K$ coupling, the relevant effective Lagrangians can be,
\begin{eqnarray}
\mathcal{L}_{\Xi^\prime Y_{1/2^\pm}K} &=& \frac{f_{ \Xi^\prime Y_{1/2^\pm}K}}{ m_{\pi}} \bar{\Xi}^{\prime \mu}
	\left\{	
	\begin{array}{c}
		1 \\ \gamma_5
	\end{array}
	\right\}
	Y_{1/2^\pm} \partial_{\mu} K+H.c,\nonumber\\
\mathcal{L}_{\Xi^\prime Y_{3/2^\pm} K}&=& (\mp i) f_{ \Xi^\prime Y_{3/2^\pm}K} \bar{\Xi}^{\prime \mu} 
	\left\{
	\begin{array}{c}
		\gamma_5 \\ 1
	\end{array}
	\right\}
	Y_{3/2^\pm}^{\mu} K+H.c,\nonumber \\
\mathcal{L}_{\Xi^\prime Y_{5/2^\pm} K}&=&  \frac{f_{ \Xi^\prime Y_{5/2^\pm}K}}{{ m_{\pi}}} \bar{\Xi}^{\prime \mu}
	\left\{
	\begin{array}{c}
		1 \\ \gamma_5
	\end{array}
	\right\}
	Y_{5/2^\pm}^{\mu \nu}\partial_{\nu}K+H.c, \quad
\label{Eq.4}
\end{eqnarray}
respectively. Hereafter the $\Xi^\prime$  refers  to $\Xi(1530)$. The details of the involved $\Lambda$ and $\Sigma$ resonances will be discussed in the following sections. With the above effective Lagrangians, one can obtain the amplitudes of the $K^- p \to K^{+} \Xi(1530)^{-}$ process, which are,
\begin{widetext}
	\begin{eqnarray}
		\mathcal{M}^{s}_{Y_{1/2^+}}&=&\bar{u}_{\mu}(p_4,m_4)\left[\frac{f_{\Xi^\prime Y_{1/2^+} K }}{{ m_{\pi}}}(i p^{\mu}_{3}) \right ] \left[S^{1/2}(k_1,m_{Y}, \Gamma_{Y})\right ] \left[-i f_{p Y_{1/2^+} K } \gamma_{5} \right ] u(p_2,m_2) F(q^{2}_{i},\Lambda^{2}_{a}) F(q^{2} _{f},\Lambda^{2}_{a}),\nonumber\\
		\mathcal{M}^{u}_{Y_{1/2^+}}&=&\bar{u}_{\mu}(p_4,m_4)\left[\frac{f_{\Xi^\prime Y_{1/2^+} K }}{{ m_{\pi}}}(-i p^{\mu}_{1}) \right ] \left[S^{1/2}(k_1,m_{Y},\Gamma_{Y})\right] \left[-i f_{p Y_{1/2^+} K} \gamma_{5} \right ] u(p_2,m_2) F(q^{2}_{i},\Lambda^{2}_{a}) F(q^{2} _{f},\Lambda^{2}_{a}),\nonumber\\
		\mathcal{M}^{s}_{Y_{1/2^-}}&=&\bar{u}_{\mu}(p_4,m_4)\left[\frac{f_{\Xi^\prime Y_{1/2^-} K }}{{ m_{\pi}}} \gamma_{5}(i p^{\mu}_{3}) \right ] \left[S^{1/2}(k_1,m_{Y},\Gamma_{Y})\right] \left[i f_{p Y_{1/2^-} K} \right ] u(p_2,m_2) F(q^{2}_{i},\Lambda^{2}_{a}) F(q^{2} _{f},\Lambda^{2}_{a}),\nonumber\\
		\mathcal{M}^{u}_{Y_{1/2^-}}&=&\bar{u}_{\mu}(p_4,m_4)\left[\frac{f_{\Xi^\prime Y_{1/2^-} K }}{{ m_{\pi}}} \gamma_{5} (-i p^{\mu}_{1}) \right ] \left[S^{1/2}(k_1,m_{Y},\Gamma_{Y})\right] \left[i f_{p Y_{1/2^-} K} \right ] u(p_2,m_2) F(q^{2}_{i},\Lambda^{2}_{a}) F(q^{2} _{f},\Lambda^{2}_{a}),\nonumber\\
		\mathcal{M}^{s}_{Y_{3/2^+}}&=&\bar{u}_{\mu}(p_4,m_4)\left[-i f_{\Xi^\prime Y_{3/2^+} K} \gamma_{5}\right]\left[S^{3/2}_{\mu \nu}(k_1,m_{Y},\Gamma_{Y})\right] \left[\frac{f_{p Y_{3/2^+} K}}{{m_{\pi}}} (-ip^{\nu}_{1})\right] u(p_2,m_2)F(q^{2}_{i},\Lambda^{2}_{a})F(q^{2}_{f},\Lambda^{2}_{a}),\nonumber\\
		\mathcal{M}^{u}_{Y_{3/2^+}}&=&\bar{u}_{\mu}(p_4,m_4)\left[-i f_{\Xi^\prime Y_{3/2^+} K}\gamma_{5} \right]\left[S^{3/2}_{\mu \nu}(k_1,m_{Y},\Gamma_{Y})\right] \left[\frac{f_{p Y_{3/2^+} K}}{{m_{\pi}}} (ip^{\nu}_{3})\right] u(p_2,m_2)F(q^{2}_{i},\Lambda^{2}_{a})F(q^{2}_{f},\Lambda^{2}_{a}),\nonumber\\
		\mathcal{M}^{s}_{Y_{3/2^-}}&=&\bar{u}_{\mu}(p_4,m_4)\left[i f_{\Xi^\prime Y_{3/2^-} K} \right]\left[S^{3/2}_{\mu \nu}(k_1,m_{Y},\Gamma_{Y})\right] \left[\frac{f_{p Y_{3/2^-} K}}{{m_{\pi}}} \gamma_{5}
		(-ip^{\nu}_{1})\right]u(p_2,m_2) F(q^{2}_{i},\Lambda^{2}_{a}) F(q^{2} _{f},\Lambda^{2}_{a}),\nonumber\\
		\mathcal{M}^{u}_{Y_{3/2^-}}&=&\bar{u}_{\mu}(p_4,m_4)\left[i f_{\Xi^\prime Y_{3/2^-} K} \right]\left[S^{3/2}_{\mu \nu}(k_1,m_{Y},\Gamma_{Y})\right] \left[\frac{f_{p Y_{3/2^-} K}}{{m_{\pi}}} \gamma_{5}
		(ip^{\nu}_{3})\right ] u(p_2,m_2) F(q^{2}_{i},\Lambda^{2}_{a}) F(q^{2} _{f},\Lambda^{2}_{a}),\nonumber\\
		\mathcal{M}^{s}_{Y_{5/2^+}}&=&\bar{u}_{\mu_{1}}(p_4,m_4)\left[\frac{f_{\Xi^\prime Y_{5/2^+} K }}{{ m_{\pi}}} (i p^{\mu_{2}}_{3}) \right]\left[S^{5/2}_{\mu_{1} \nu_{1} \mu_{2} \nu_{2}}(k_1,m_{Y},\Gamma_{Y})\right] \left[-i \frac{f_{p Y_{5/2^+} K}}{{m^{2}_{\pi}}} \gamma_{5} (-p^{\nu_{1}}_{1} p^{\nu_{2}}_{1})\right] u(p_2,m_2)F(q^{2}_{i},\Lambda^{2}_{b})F(q^{2}_{f},\Lambda^{2}_{b}),\nonumber\\
		\mathcal{M}^{u}_{Y_{5/2^+}}&=&\bar{u}_{\mu_{1}}(p_4,m_4)\left[\frac{f_{\Xi^\prime Y_{5/2^+} K }}{{ m_{\pi}}} (-i p^{\mu_{2}}_{1}) \right]\left[S^{5/2}_{\mu_{1} \nu_{1} \mu_{2} \nu_{2}}(k_1,m_{Y},\Gamma_{Y})\right] \left[-i \frac{f_{p Y_{5/2^+} K}}{{m^{2}_{\pi}}} \gamma_{5} (-p^{\nu_{1}}_{3} p^{\nu_{2}}_{3})\right] u(p_2,m_2)F(q^{2}_{i},\Lambda^{2}_{b})F(q^{2}_{f},\Lambda^{2}_{b}),\nonumber\\
		\mathcal{M}^{s}_{Y_{5/2^-}}&=&\bar{u}_{\mu_{1}}(p_4,m_4)\left[\frac{f_{\Xi^\prime Y_{5/2^-} K }}{{ m_{\pi}}} \gamma_{5} (i p^{\mu_{2}}_{3}) \right]\left[S^{5/2}_{\mu_{1} \nu_{1} \mu_{2} \nu_{2}}(k_1,m_{Y},\Gamma_{Y})\right] \left[-i \frac{f_{p Y_{5/2^-} K}}{{m^{2}_{\pi}}} (-p^{\nu_{1}}_{1} p^{\nu_{2}}_{1})\right] u(p_2,m_2)F(q^{2}_{i},\Lambda^{2}_{b})F(q^{2}_{f},\Lambda^{2}_{b}),\nonumber\\
		\mathcal{M}^{u}_{Y_{5/2^-}}&=&\bar{u}_{\mu_{1}}(p_4,m_4)\left[\frac{f_{\Xi^\prime Y_{5/2^-} K }}{{ m_{\pi}}} \gamma_{5} (-i p^{\mu_{2}}_{1}) \right]\left[S^{5/2}_{\mu_{1} \nu_{1} \mu_{2} \nu_{2}}(k_1,m_{Y},\Gamma_{Y})\right] \left[-i \frac{f_{p Y_{5/2^-} K}}{{m^{2}_{\pi}}} (-p^{\nu_{1}}_{3} p^{\nu_{2}}_{3})\right] u(p_2,m_2)F(q^{2}_{i},\Lambda^{2}_{b})F(q^{2}_{f},\Lambda^{2}_{b}),\nonumber\\
        \label{Eq:Amp1}
\end{eqnarray}
\end{widetext}
where the superscripts $s$ and $u$ correspond to $s$ and $u$ channels, respectively, and the subscript $Y_{J^{P}}$ corresponds to the hyperon involved in the processes. In the above amplitudes, the $\mathcal{S}^{1/2}(k_i,m_i,\Gamma_i)$ refers to the propagator corresponding to the baryon with spin $1/2$, which is~\cite{Jackson:2015dva},
\begin{eqnarray}
	\mathcal{S}^{1/2}(k_i,m_i,\Gamma_i) =\frac{\slash\!\!\!k_{i}+m_{i}} {k^2_{i}-m^2_{i} +i m_i \Gamma_{i}},
\end{eqnarray}
and $\mathcal{S}^{3/2}_{\mu \nu}(k_i,m_i,\Gamma_i)$ is the propagator corresponding to the baryon with spin $3/2$, which reads,
\begin{eqnarray}
		\mathcal{S}^{3/2}_{\mu \nu}(k_i,m_i,\Gamma_i) &=&\frac{\slash\!\!\!k_{i}+m_{i}} {k^2_{i}-m^2_{i} +i m_i \Gamma_{i}}\nonumber \Big(-g^{\mu \nu} +\frac{1}{3}\gamma^{\mu}\gamma^{\nu}\\&+&\frac{2p^{\mu} p^{\nu}}{3m^2_{i}}+\frac{\gamma^{\mu}p^{\nu}-p^{\mu}\gamma^{\nu}}{3m_{i}}\Big).
\end{eqnarray}
The $\mathcal{S}^{5/2}_{\mu_{1} \mu_{2} \nu_{1} \nu_{2}}(k_i,m_i,\Gamma_i)$ is the propagator corresponding to the baryon with spin $5/2$, and its detailed expression is ~\cite{David:1995pi},
\begin{eqnarray}
\mathcal{S}^{5/2}_{\mu_{1} \mu_{2} \nu_{1} \nu_{2}}(k_i,m_i,\Gamma_i) &=&\frac{\slash\!\!\!k_{i}+\sqrt{s}} {10(k^2_{i}-m^2_{i} +i m_i \Gamma_{i})}\nonumber\\ &\times& \Big(5\tilde{g}_{\mu_{1} \mu_{2}} \tilde{g}_{\nu_{1} \nu_{2}}-2 \tilde{g}_{\mu_{1} \nu_{1}} \tilde{g}_{\mu_{2} \nu_{2}}\nonumber\\ &+&5 \tilde{g}_{\mu_{1} \nu_{2}} \tilde{g}_{\nu_{1} \mu_{2}} +\tilde{g}_{\mu_{1} \rho}\gamma^{\rho} \gamma^{\sigma} \tilde{g}_{\sigma \mu_{2}} \tilde{g}_{\nu_{1} \nu_{2}}\nonumber\\ &+& \tilde{g}_{\nu_{1} \rho}\gamma^{\rho} \gamma^{\sigma} \tilde{g}_{\sigma \nu_{2}} \tilde{g}_{\mu_{1} \mu_{2}} + \tilde{g}_{\mu_{1} \rho} \gamma^{\rho} \gamma^{\sigma} \tilde{g}_{\sigma \nu_{2}} \tilde{g}_{\nu_{1} \mu_{2}}\nonumber\\ &+& \tilde{g}_{\nu_{1} \rho} \gamma^{\rho} \gamma^{\sigma} \tilde{g}_{\rho \mu_{2}} \tilde{g}_{\mu_{1} \nu_{2}}\Big),
\end{eqnarray}
with
\begin{eqnarray}
\tilde{g}_{\mu \nu}=-g_{\mu \nu}+ {q_{\mu} q_{\nu}}/{s}.
\end{eqnarray}

In addition, two form factors are introduced to depict the two relevant vertices, and  both form factors are in the form, 
\begin{eqnarray}
  F({\vec{q}}^{2},\Lambda^{2}_{i})=\mathrm{exp}(-{\vec{q}}^{2}/{\Lambda^{2}_{i}}),
\end{eqnarray}
where $\Lambda_{i}$ stands for the cutoff constant. In the present estimations, we adopt two different values  of $\Lambda_{i}$, which is $\Lambda_{a}$ for low-spin propagator ($J^{P}\leq 3/2$) and $\Lambda_{b}$ for high-spin propagator ($J\ge 5/2$) as in Ref.~\cite{Sharov:2011xq}. In addition, $\vec{q_{i}}$ and $\vec{q_{f}}$ stand for three momenta of the initial $K^{-}$ and the final $K^{+}$ in the center-of-mass system, respectively, which read,
\begin{eqnarray}
{\vec{q}}_{i} &=&\sqrt{\frac{(s-(m_{1}-m_{2})^{2})(s-(m_{1}+m_{2})^{2})}{4s}}, \nonumber\\
{\vec{q}}_{f} &=&\sqrt{\frac{(s-(m_{3}-m_{4})^{2})(s-(m_{3}+m_{4})^{2})}{4s}}.\label{Eq:qiqf}
\end{eqnarray}

\begin{figure}[t]
	\begin{tabular}{ccc}
		\centering
		\includegraphics[width=75mm]{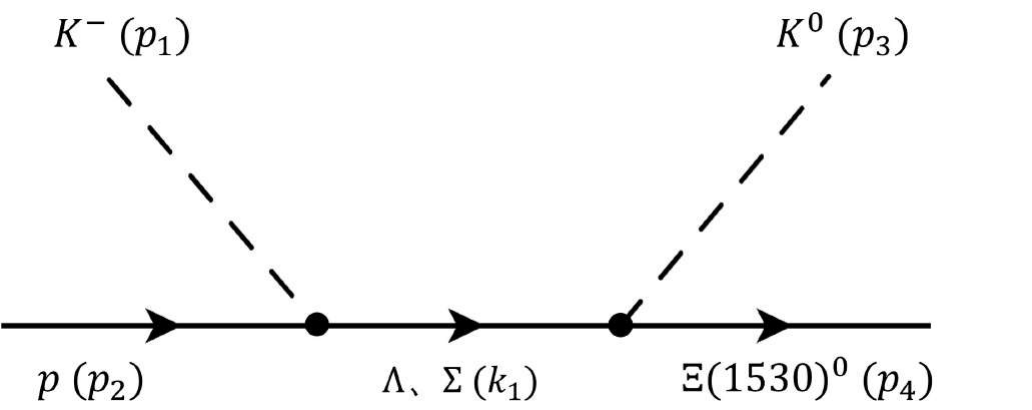}&\\
		{\large(a)} &\\
		\includegraphics[width=75mm]{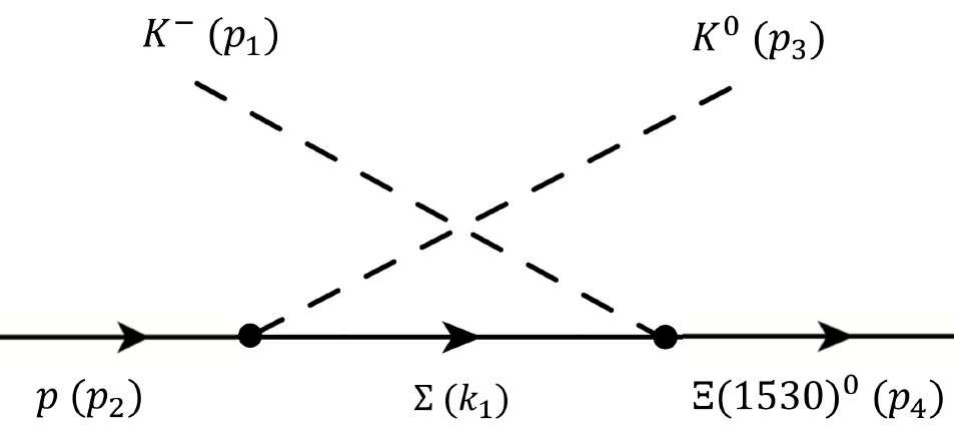}&\\
		\\
		\large {(b)}& \\
	\end{tabular}
	\caption{Diagrams contributing to the process of $K^{-}p \rightarrow K^{0} \Xi(1530)^{0}$, 
		 corresponding to the (a) $s$- and (b) $u$-channel contributions, respectively.}\label{Fig.2}
\end{figure}

In the same way, one can investigate the $\Xi(1530)^{0}$ production in the $K^{-} p$ scattering process, and the relevant diagrams are shown in Fig.~\ref{Fig.2}. It should be noted that in the $u$ channel, the exchanged hyperons could only be $\Sigma$ and its resonances, while the $\Lambda$ hyperon and its resonances are forbidden due to charge conservation. Thus, from the isospin symmetry, the contributions from the $\Sigma$ hyperon and its resonances are the same in both $K^- p \to K^+ \Xi(1530)^-$ and $K^- p \to K^0 \Xi(1530)^0$ processes, while those from the $\Lambda$ hyperon and its resonances are different in both processes.

\section{NUMERICAL RESULTS AND DISCUSSIONS}
\label{sec:MA}

\begin{table}
	\caption{The resonance parameters of the involved $\Lambda$ and $\Sigma$ hyperons and their resonances.}
	\label{Tab.1}
\renewcommand{\arraystretch}{2}
	\setlength{\tabcolsep}{12pt}
	\centering
	\begin{tabular}{cccc}
		\hline\hline
		State&$J^{P}$&Mass (MeV)& Width (MeV)\\
		\hline
		$\Lambda(1116)$&$1/2^{+}$&1116&--\\
		$\Lambda(1405)$&$1/2^{-}$&1405&50\\
		$\Lambda(1890)$&$3/2^{+}$&1890&120\\
		$\Lambda(2110)$&$5/2^{+}$&2090&250\\
		$\Lambda(2325)$&$3/2^{-}$&2325&170\\
		$\Sigma(1193)$&$1/2^{+}$&1193&--\\
		$\Sigma(1670)$&$3/2^{-}$&1662&55\\
		$\Sigma(2080)$&$3/2^{+}$&2090&170\\
		$\Sigma(2250)$&$5/2^{-}$&2250&100\\
		\hline		
	\end{tabular}
\end{table}

It is worth mentioning that a lot of $\Lambda$ and $\Sigma$ resonances have been observed below the threshold of $K \Xi(1530)$, and these states dominantly contribute to the $u$ channel of the reaction process, and the lineshape of the cross section resulted from these below-threshold states are very similar, thus, it is very difficult to determine the individual contributions from these below-threshold states by fitting the cross sections for $K^- p \to K\Xi(1530)$. In the present fit, we only consider $\Lambda(1116)$, $\Lambda(1405)$, $\Lambda(1890)$, $\Sigma(1193)$ and $\Sigma(1670)$, whose $J^P$ quantum numbers are $1/2^{\pm}$ or $3/2^\pm$. As for the above-threshold states, the contributions from the $s$ channel are dominant, by comparing the structure of the cross sections~\cite{Berge:1966zz, Dauber:1969hg, Briefel:1977bp, Flaminio:1979iz}, we take $\Sigma(2080)$, $\Lambda(2110)$, $\Sigma(2250)$, and $\Lambda(2325)$ into consideration. The resonance parameters involved in the present fit are collected in Table~\ref{Tab.1}. As indicated in Ref.~\cite{Sharov:2011xq}, for the above-threshold resonance with high-spin, the $s$-channels are dominant compared to the $u$-channel, thus, in the present estimations, only $s$-channels are considered for $\Lambda(2110)$ and $\Sigma(2250)$. Therefore, the amplitude for $K^- p \to K^+ \Xi(1530)^-$ reads,
\begin{eqnarray}
\mathcal{M}^{\mathrm{Tot}}_{K^+\Xi^{\prime -}}&=&\mathcal{M}^{s}_{\Lambda(1116)}+ \mathcal{M}^{u}_{\Lambda(1116)}+\mathcal{M}^{s}_{\Sigma(1193)}+ \mathcal{M}^{u}_{\Sigma(1193)}\nonumber\\  &+&\mathcal{M}^{s}_{\Lambda(1405)}+ \mathcal{M}^{u}_{\Lambda(1405)} +\mathcal{M}^{s}_{\Sigma(1670)}+ \mathcal{M}^{u}_{\Sigma(1670)}\nonumber\\
&+&\mathcal{M}^{s}_{\Lambda(1890)}+ \mathcal{M}^{u}_{\Lambda(1890)} +\mathcal{M}^{s}_{\Sigma(2080)}+\mathcal{M}^{s}_{\Sigma(2080)}\nonumber\\&+&\mathcal{M}^{s}_{\Lambda(2110)}+ \mathcal{M}^{u}_{\Lambda(2110)} +\mathcal{M}^{s}_{\Sigma(2250)}+\mathcal{M}^{s}_{\Sigma(2250)}\nonumber\\ &+&\mathcal{M}^{s}_{\Lambda(2325)}+ \mathcal{M}^{u}_{\Lambda(2325)}. \quad
\label{Eq:A1}
\end{eqnarray}
Similarly, one can obtain the amplitudes for $K^- p \to K^0 \Xi(1530)^0$, which is,
\begin{eqnarray}
\mathcal{M}^{\mathrm{Tot}}_{K^0\Xi^{\prime 0}}&=&\mathcal{M}^{s}_{\Lambda(1116)}+\mathcal{M}^{s}_{\Sigma(1193)}+ \mathcal{M}^{u}_{\Sigma(1193)}+\mathcal{M}^{s}_{\Lambda(1405)}\nonumber\\&+&\mathcal{M}^{s}_{\Sigma(1670)}+ \mathcal{M}^{u}_{\Sigma(1670)}+\mathcal{M}^{s}_{\Lambda(1890)}+\mathcal{M}^{s}_{\Sigma(2080)}\nonumber\\&+&\mathcal{M}^{u}_{\Sigma(2080)}+ \mathcal{M}^{s}_{\Lambda(2110)}+\mathcal{M}^{s}_{\Sigma(2250)}+\mathcal{M}^{u}_{\Sigma(2250)}\nonumber\\&+&\mathcal{M}^{s}_{\Lambda(2325)}.\quad
\label{Eq:A2}
\end{eqnarray}

With the above amplitudes, one can obtain the differential cross sections for the $K^-p \to K\Xi(1530)$ process,
\begin{eqnarray}
	\frac{d{\sigma}} {d \cos\theta}
	=\frac{1} {32 \pi s} \frac{|\vec{q}_f|} {|\vec{q}_i|}  \left(\frac{1} {2} {\bigg|\overline{\mathcal{M}^{\mathrm{Tot}}_{K\Xi^{\prime}}}\bigg|}^2\right),
\end{eqnarray}
where $s$ refers to the square of the center-of-mass energy, and $\theta$ is the scattering angle, which is the angle of outgoing $\Xi(1530)$ and the incident kaon beam direction in the  center-of-mass frame. The $\vec{q_{f}}$ and $\vec{q_{i}}$ stand for the three-momentum of the final $\Xi(1530)$ and the initial kaon beam in the center-of-mass system, respectively, and their values are presented in Eqs.~\eqref{Eq:qiqf}.

As indicated in Eqs.~\eqref{Eq:Amp1}, the coupling constants $f_{\Xi^{\prime} Y_{J^P}K}$ and $f_{pY_{J^P}K}$ always appear in the same amplitude, and the amplitude is proportional to the product of these two coupling constants. Thus, we can define the product of these two coupling constants as one single parameter, which is, 
\begin{eqnarray}
g_{Y}=f_{\Xi^{\prime} Y K} \times f_{p Y K},
\end{eqnarray}
where $Y$ refers to the $\Lambda$ and $\Sigma$ hyperons and their resonances. In the present estimations, the values of $g_{Y}$ will be determined by simultaneously fitting the cross sections for $K^-p \to K^+\Xi(1530)^-$ and $K^-p \to K^0\Xi(1530)^0$ .

\begin{figure*}[t]
     \includegraphics[width=180mm]{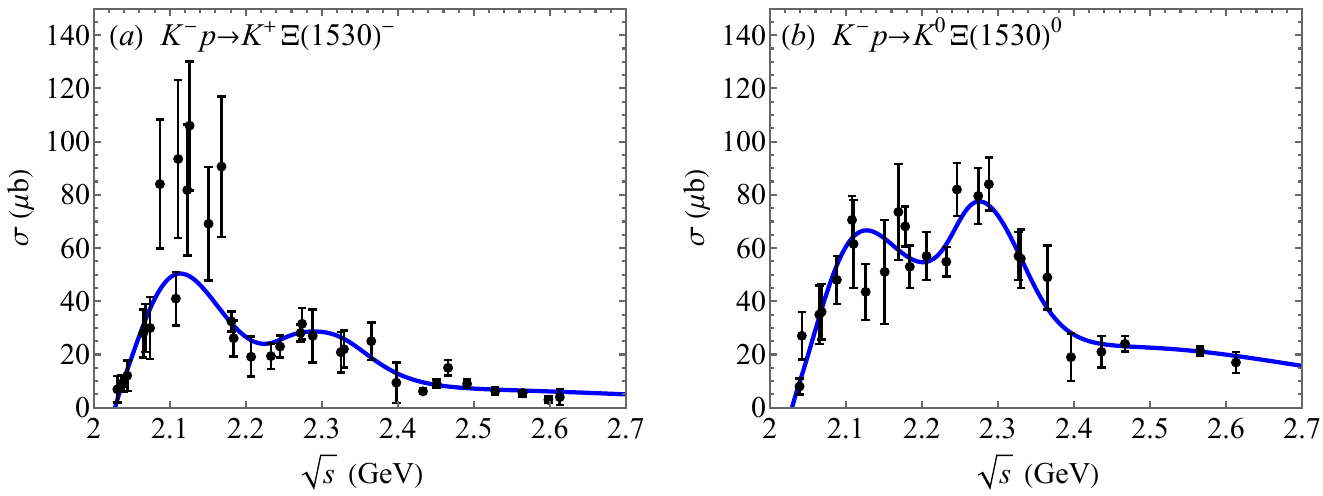}
     \caption{(Color online) The cross sections for $(a)$ $K^{-} p \rightarrow K^+ \Xi(1530)^-$ and $(b)$ $K^{-} p \rightarrow K^{0} \Xi(1530)^{0}$ depending on the $\sqrt{s}$. The black points with error bars correspond to the experimental data of the cross sections for $K^- p \to K^+ \Xi(1530)^-$ and $K^- p \to K^0 \Xi(1530)^0$ from Refs.~\cite{Berge:1966zz, Dauber:1969hg, Briefel:1977bp, Flaminio:1979iz}, while the blue solid curves are obtained with the fitted parameters in Table~\ref{Tab.2}.}
\label{Fig.3}
\end{figure*}

\begin{figure}[t]
	\centering
     \includegraphics[width=85mm]{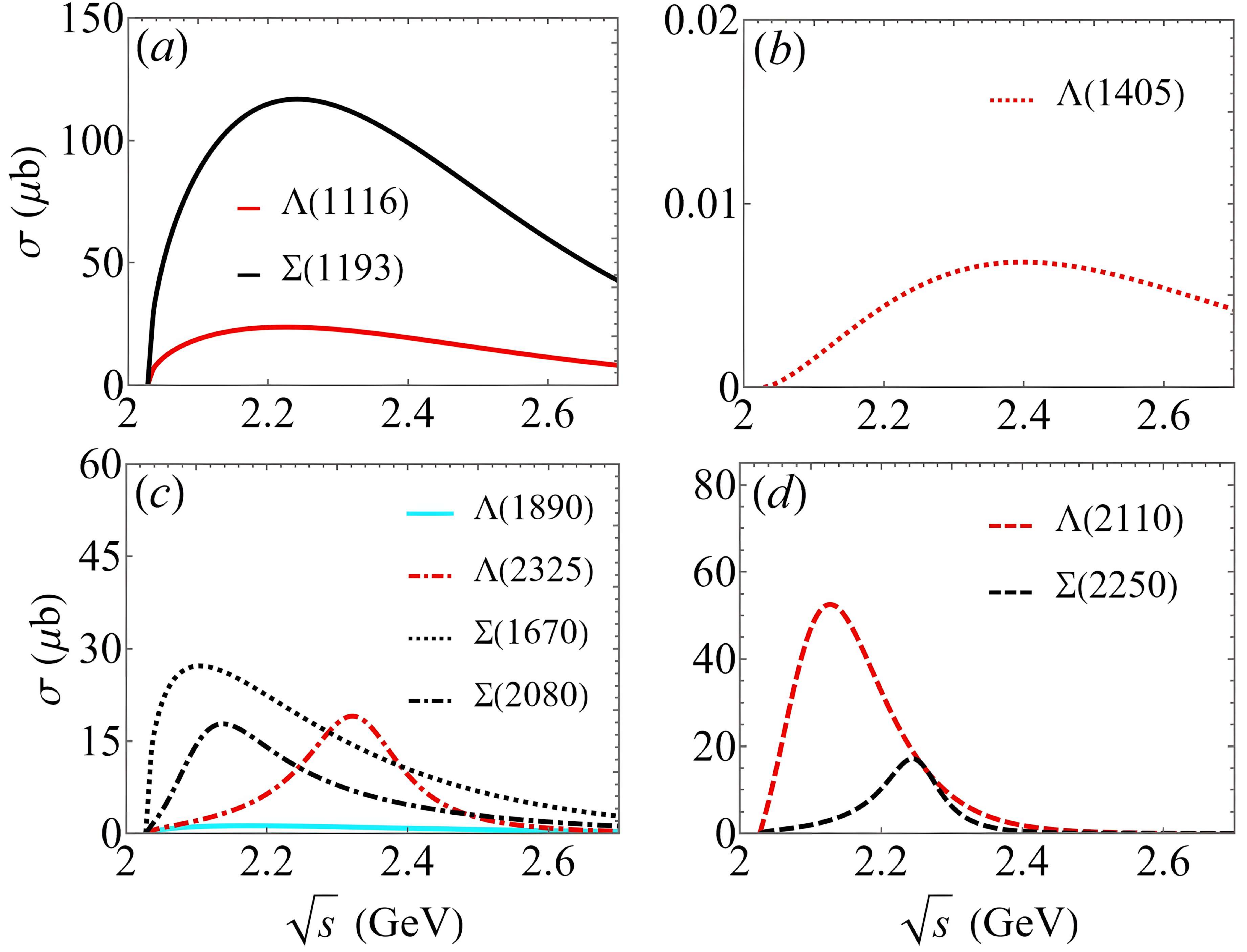}
 \caption{(Color online) The individual contributions from different intermediate states for the cross sections for  $K^{-} p \rightarrow K^{+} \Xi(1530)^{-}$ depending on the $\sqrt{s}$ in model A. } 
	\label{Fig.4}
\end{figure}

\begin{figure}[htb]
	\centering
     \includegraphics[width=85mm]{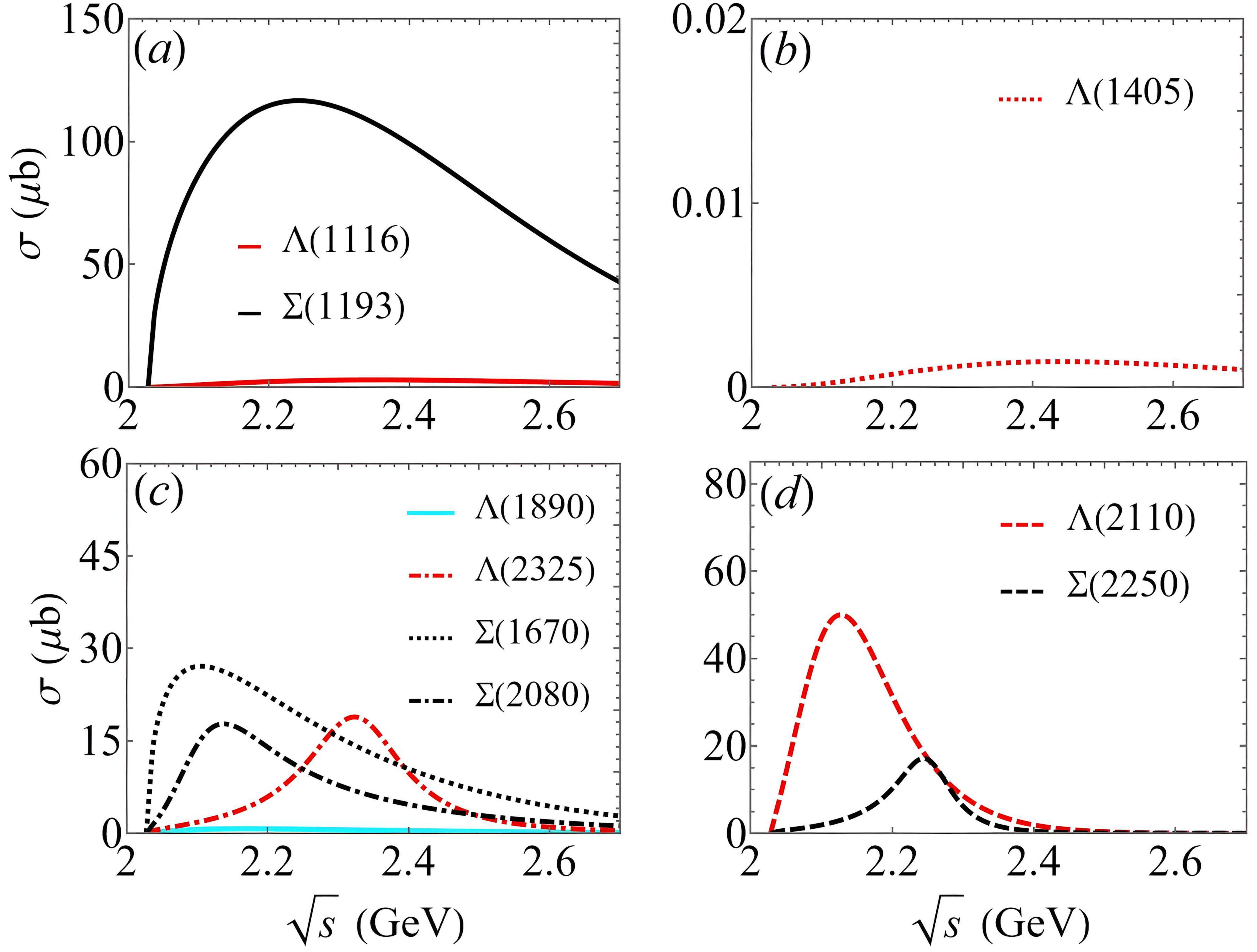}
 \caption{(Color online) The individual contributions from different intermediate states for the cross sections for  $K^{-} p \rightarrow K^{0} \Xi(1530)^{0}$ depending on the $\sqrt{s}$ in model A. It should be noted that the diagrams $(c)$ and $(d)$ are almost the same as those in Fig.~\ref{Fig.4} since the above-threshold $\Lambda$ resonance contributions are dominated by the s channel, while the $\Sigma$ resonance contributions are exactly the same in both processes.} 
	\label{Fig.5}
\end{figure}

\subsection{Cross sections for $K^- p \to K \Xi(1530)$}
In the present fitting process, nine $\Lambda$ and $\Sigma$ hyperons and their resonances are taken into consideration. In addition, two kinds of cutoff constants, $\Lambda_{a}$ for low-spin intermediate states ($J\leq 3/2$) and $\Lambda_{b}$ for high-spin intermediate states ($J\ge 5/2$), are involved by the form factors in the amplitudes. Here, we take $\Lambda_a=1.0$ GeV and $\Lambda_b=0.5$ GeV as those in Ref.~\cite{Sharov:2011xq}, while the remaining parameters $g_Y$ can be determined by fitting the cross sections for $K^- p \to K^+ \Xi(1530)^-$ and $K^- p \to K^0 \Xi(1530)^0$ simultaneously. In addition, it should be noted that the experimental data for the cross sections for $K^- p \to K^+ \Xi(1530)^-$ in the range of $\sqrt{s}=[2.087,2.168]$ GeV are inconsistent with each other as shown in Fig.~\ref{Fig.3}. In particular, the cross sections at $2.087$, $2.111$, $2.123$, $2.126$, $2.151$, and $2.168$ GeV are around 80 $\mathrm{\mathrm{\mu}b}$ with large uncertainties, while the cross section at $2.108$ GeV is measured to be $(41 \pm 10)\ \mathrm{\mu b}$, which is evidently smaller than the nearby data. Due to the inconsistency mentioned above,  we employ two distinct fitting strategies, a uniform weighting scheme (model A) and different weighting approach (model B).

\begin{table}[t]
	\caption{The fitted parameters of the process $K^{-}p \rightarrow K \Xi(1530)$, where the weight of each experimental value remains the same.}
	\label{Tab.2}
	\renewcommand{\arraystretch}{2}
	\setlength{\tabcolsep}{2pt}
	\centering
	\begin{tabular}{p{1.5cm}<\centering p{2cm}<\centering p{1.5cm}<\centering p{2cm}<\centering}
		\hline\hline
		Coupling&Value& Coupling&Value\\
		\hline
           $g_{\Lambda(1116)}$&$1.37\pm +0.20$& 
           $g_{\Sigma(1193)}$&$-3.18\pm 0.07$\\
		$g_{\Lambda(1405)}$&$-0.05\pm 0.08$&
		$g_{\Sigma(1670)}$&$-1.55\pm 0.20$\\
		$g_{\Lambda(1890)}$&$0.50\pm 0.27$&
		$g_{\Sigma(2080)}$&$-0.86\pm 0.20$\\
		$g_{\Lambda(2110)}$&$0.45\pm 0.02$&
		$g_{\Sigma(2250)}$&$0.21\pm 0.03$\\
		$g_{\Lambda(2325)}$&$-0.18\pm 0.02$\\
		\multicolumn{4}{c}{$\chi^{2}= 63.38$}\\
		\hline		
	\end{tabular}	
\end{table}

\begin{figure*}[htb]
     \includegraphics[width=180mm]{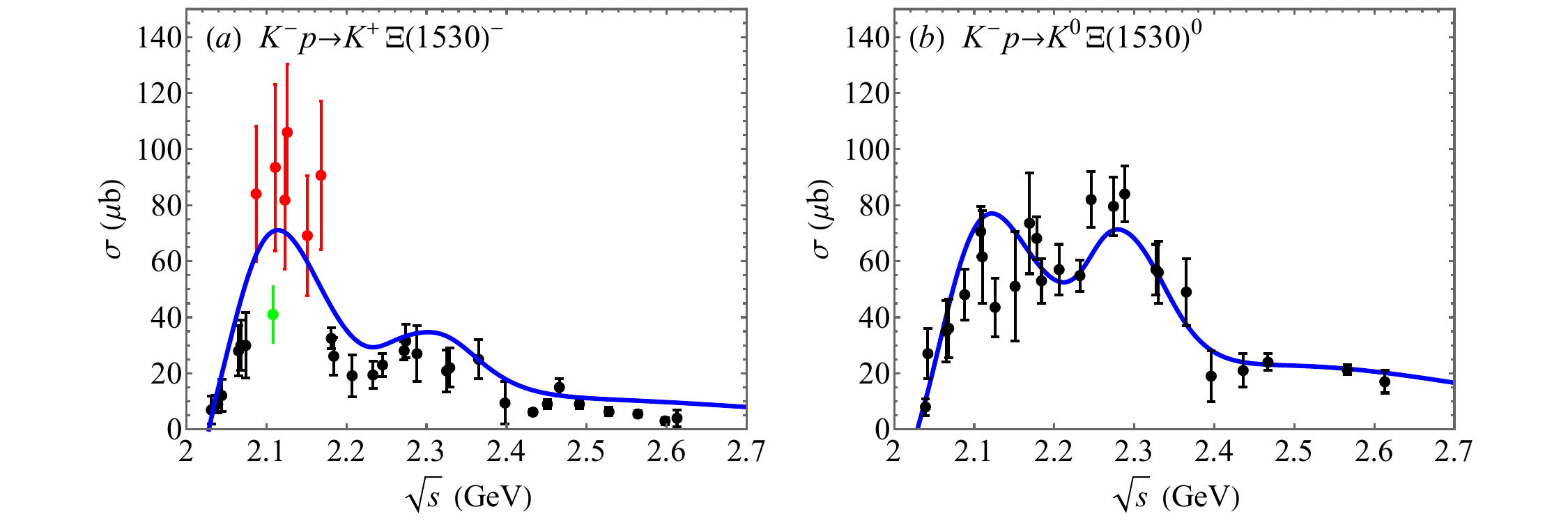}
     \caption{(Color online) The same as Fig.~\ref{Fig.3} but in model B. The points with error bars correspond to the experimental data of the cross sections for $K^- p \to K^+ \Xi(1530)^-$ and $K^- p \to K^0 \Xi(1530)^0$ from Refs.~\cite{Berge:1966zz, Dauber:1969hg, Briefel:1977bp, Flaminio:1979iz}, while the blue solid curves are obtained with the fitted parameters in Table~\ref{Tab.3}. The green data are excluded in the present fit, while the weight of the six red data points in the range $\sqrt{s}=[2.087,2.168]$ GeV are set to  8.}
\label{Fig.6}
\end{figure*}

\subsubsection{Model A}
In this scenario, the parameters $g_{Y}$ determined by fitting the cross sections for $K^- p \to K^+ \Xi(1530)^-$ and $K^- p \to K^0 \Xi(1530)^0$ are collected in Table~\ref{Tab.2}. With the central values of the determined parameters, one has $\chi^2 /\mathrm{DOF}= 63.38/50=1.27$, which is acceptable for the present fit. In Fig.~\ref{Fig.3}, we present the fitted cross sections for $K^- p \to K^+ \Xi(1530)^-$ [Fig~\ref{Fig.3}$(a)$] and $K^- p \to K^0 \Xi(1530)^0$ [Fig~\ref{Fig.3}$(b)$], where the experimental data from Refs.~\cite{Berge:1966zz, Dauber:1969hg, Briefel:1977bp, Flaminio:1979iz} are also presented for comparison. From the figure one can find that the cross sections for $K^- p \to K^0 \Xi(1530)^0$ [Fig~\ref{Fig.3}$(b)$] can be well reproduced in the whole considered range.  
Meanwhile, the cross sections for $K^- p \to K^0 \Xi(1530)^0$ can also be reproduced except for the data in the range $\sqrt{s}=[2.087,2.168]$ GeV due to  the inconsistency of the experimental data. Specifically, the cross section for $K^- p \to K^+ \Xi(1530)^-$ is fitted to be 50.41 $\mu$b at $\sqrt{s}=2.11$ GeV, which is consistent with the experimental measurement within uncertainty. However, the fitted cross sections in the range $\sqrt{s}=[2.087,2.168]$ GeV are overall smaller than other data in this range. Furthermore, it is worth noting that there are two clear peak structures around $\sqrt{s}=2.1$ GeV and $\sqrt{s}=2.3$ GeV in both cross sections.

\begin{figure}[t]
	\centering
     \includegraphics[width=85mm]{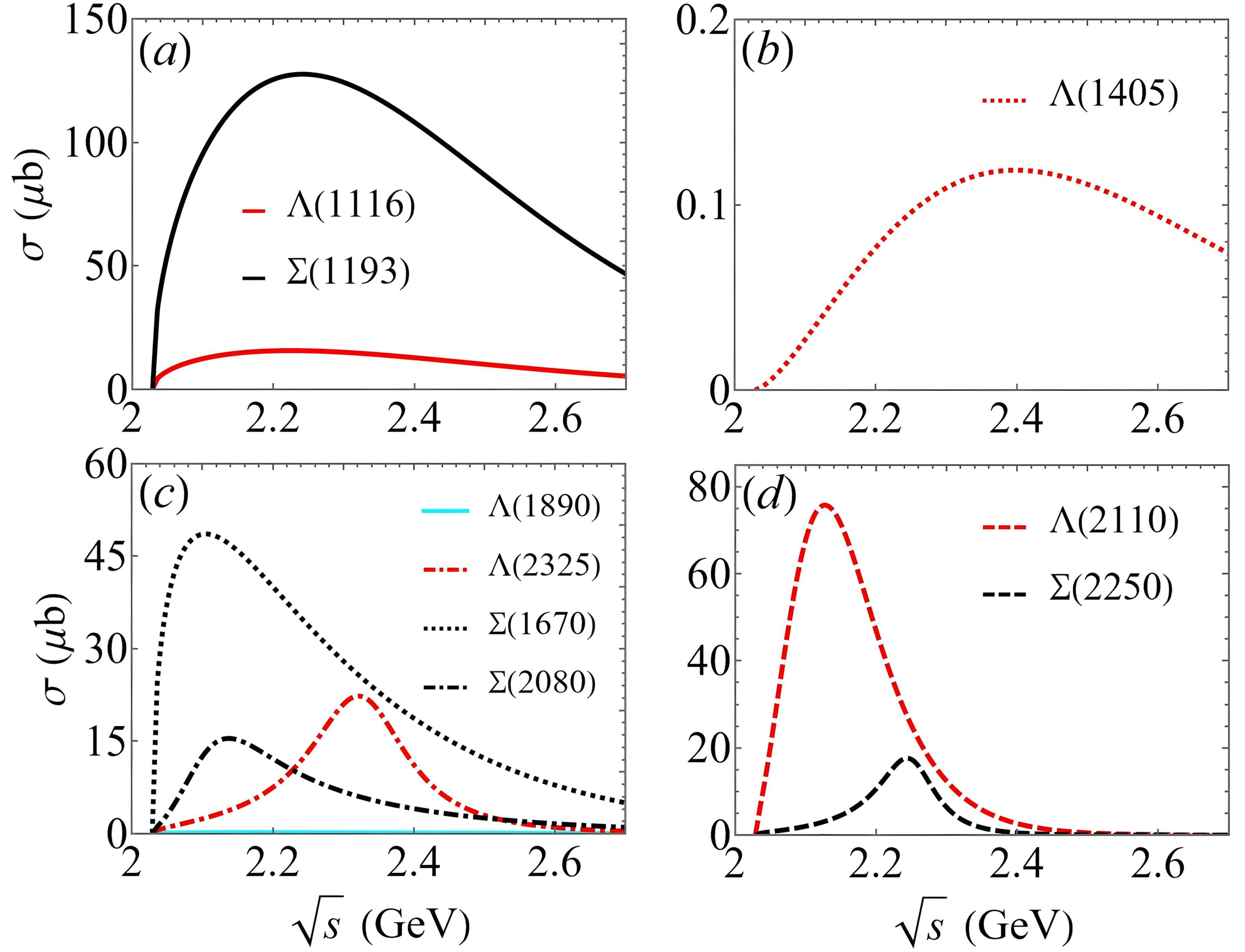}
 \caption{(Color online) The same as Fig.~\ref{Fig.4} but in model B.} 
	\label{Fig.7}
\end{figure}

\begin{figure}[t]
	\centering
     \includegraphics[width=85mm]{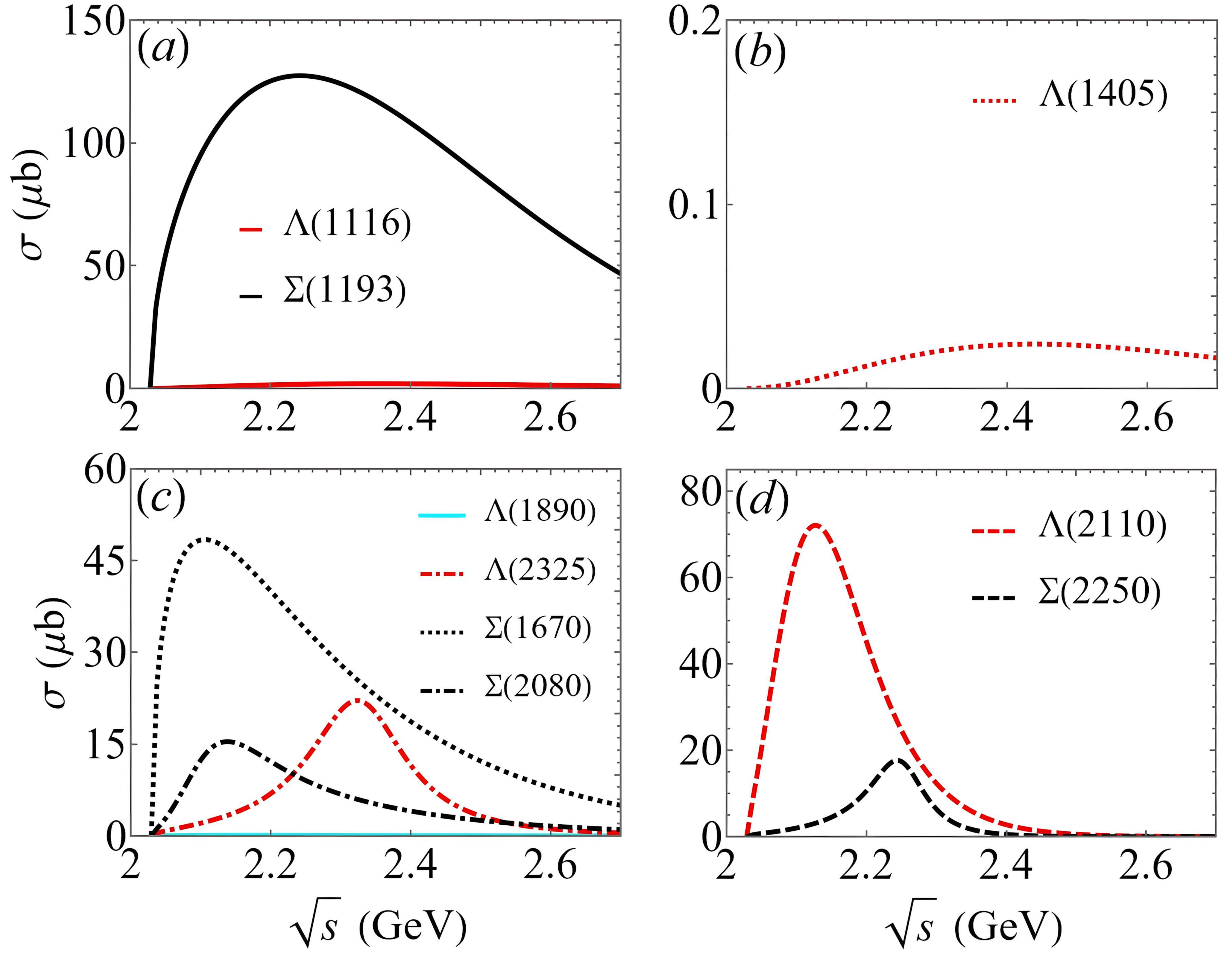}
 \caption{(Color online) The same as Fig.~\ref{Fig.5} but in model B. The diagrams $(c)$ and $(d)$ are almost the same as those in Fig.~\ref{Fig.7} for the same reason clarified in Fig.~\ref{Fig.5}.} 
	\label{Fig.8}
\end{figure}

\begin{table}
	\caption{The fitted parameters of the process $K^{-}p \rightarrow K \Xi(1530)$, where the weight of six experimental data values with high cross sections was increased (eight times), and the one experimental data value with low cross section in the range $\sqrt{s}=[2.087,2.168]$ GeV was deleted.}
	\label{Tab.3}
	\renewcommand{\arraystretch}{2}
	\setlength{\tabcolsep}{5pt}
	\centering
	\begin{tabular}{p{1.5cm}<\centering p{2cm}<\centering p{1.5cm}<\centering p{2cm}<\centering}
		\hline\hline
		Coupling&Value& Coupling&Value\\
		\hline
           $g^{\prime}_{\Lambda(1116)}$&$1.12\pm +0.22$& 
           $g^{\prime}_{\Sigma(1193)}$&$-3.32\pm 0.15$\\
		$g^{\prime}_{\Lambda(1405)}$&$-0.19\pm 0.09$&
		$g^{\prime}_{\Sigma(1670)}$&$-2.07\pm 0.28$\\
		$g^{\prime}_{\Lambda(1890)}$&$0.19\pm 0.60$&
		$g^{\prime}_{\Sigma(2080)}$&$-0.81\pm 0.22$\\
		$g^{\prime}_{\Lambda(2110)}$&$0.55\pm 0.02$&
		$g^{\prime}_{\Sigma(2250)}$&$0.21\pm 0.03$\\
		$g^{\prime}_{\Lambda(2325)}$&$-0.20\pm 0.02$\\
		\multicolumn{4}{c}{$\chi^{2}_{0}= 96.29$ \quad $\chi^{2}_{1}= 138.93$}\\
		\hline		
	\end{tabular}	
\end{table}

In addition to the total cross sections, the individual contributions of each intermediate state for $K^{-} p \rightarrow K^{+} \Xi(1530)^{-}$ are presented in Fig.~\ref{Fig.4}. From Fig.~\ref{Fig.4}($a$) one can find that the cross sections resulted from the $\Sigma(1193)$ intermediate process are predominant, while the cross section resulted from $\Lambda(1116)$ is several times smaller than that of $\Sigma(1193)$. As indicated by Fig.~\ref{Fig.4}($b$), the cross sections resulted from $\Lambda(1405)$ are at least three orders smaller than those from $\Lambda(1116)$, which indicates that the contributions from $\Lambda(1405)$ are negligible even if one considers the interference contributions with other states. In Fig.~\ref{Fig.4}($c$), we present the contributions from $\Lambda(1890)$, $\Lambda(2325)$, $\Sigma(1670)$, and $\Sigma(2080)$, whose total angular momenta are $3/2$. From this diagram, one can find that the contribution from $\Lambda(1890)$ is not obvious with its contribution to the cross sections to be less than 2 $\mu$b.  Meanwhile,  the peak values of cross sections for $\Lambda(2325)$, $\Sigma(1670)$, and $\Sigma(2080)$ are about 20 $\mu$b. In Fig.~\ref{Fig.4}($d$), we present the individual contributions from two high-spin intermediate state,  i.e. $\Lambda(2110)$ and $\Sigma(2250)$. The maximum of the cross sections resulted from $\Lambda(2110)$ is estimated to be around 50 $\mu$b at $\sqrt{s}=2.13$ GeV, and the contribution from $\Sigma(2250)$ is about 1/3 of that from $\Lambda(2110)$ in the peak position.

In addition, the individual contributions from hyperons for $K^{-} p \rightarrow K^{0} \Xi(1530)^{0}$ are presented in Fig.~\ref{Fig.5}. Comparing with Fig.~\ref{Fig.4}, one can find that the cross sections resulted from four $\Sigma$ intermediate states are basically consistent in these two processes, which are consistent with isospin symmetry expectation. As for the $\Lambda$ intermediate process, the $u$-channel contributions vanish for the $K^- p \to K^0 \Xi(1530)^0$ process due to charge conservation. Comparing Fig.~\ref{Fig.4} and Fig.~\ref{Fig.5}, one can find that the contributions from $\Lambda(1116)$, $\Lambda(1405)$, and $\Lambda(1890)$ are much larger in the $K^- p \to K^+ \Xi(1530)^-$ process, which indicates that the $u$ channel is dominant for the below-threshold intermediate states. Moreover, the contributions from the above-threshold $\Lambda$ resonances, such as $\Lambda(2110)$ and $\Lambda(2325)$ are similar, indicating that the $s$ channel is dominant.

\subsubsection{Model B}
As we clarified at the beginning of this section, the experimental data for the cross sections for $K^- p \to K^+ \Xi(1530)^-$ in the range $\sqrt{s}=[2.087,2.168]$ GeV are inconsistent with each other. In model B, we first remove the data at $\sqrt{s}=2.11$ GeV and perform a weighted fit with the weights of the other data in this range enhanced. We increase the weight of these data, and find that when we set the weight of these data to be 8, the lineshape of the cross sections for $K^- p \to K^+ \Xi(1530)^-$  can roughly reproduce the data in the range $\sqrt{s}=[2.087,2.168]$ GeV. In this case the fitted parameters are collected in Table~\ref{Tab.3}. With these parameters, the $\chi^2$ and weighted $\chi^2$ are estimated to be 96.24 and 138.93, respectively, which are much larger than those  in model A.

Using the parameters in Table~\ref{Tab.3}, we can obtain the cross sections for $K^- p \to K^+ \Xi(1530)^-$ and $K^- p \to K^0 \Xi(1530)^0$, which are presented in Fig.~\ref{Fig.6}. Notably, the cross sections for $K^- p \to K^+ \Xi(1530)^-$ approach 70 $\mathrm{\mu b}$ in the vicinity of 2.1 GeV, which is consistent with the lower limit of the experimental data. Nonetheless, the fitted cross sections overestimate the experimental measurements outside the range $\sqrt{s}=[2.087,2.168]$ GeV. For the $K^- p \to K^0 \Xi(1530)^0$ process, the fitted curve exhibits a modest overestimation of the experimental data in the vicinity of 2.1 GeV, followed by an underestimation around 2.3 GeV. Based on a comparative analysis of Figs.~\ref{Fig.4} and \ref{Fig.6}, we conclude that model A demonstrates a better global agreement with the experimental data than model B. More accurate measurements of the cross sections in this energy range can be helpful for our understanding of $\Xi(1530)$ production in the $K^- p$ reaction , which should be accessible for J-PARC.

\begin{figure*}[t]
     \includegraphics[width=175mm]{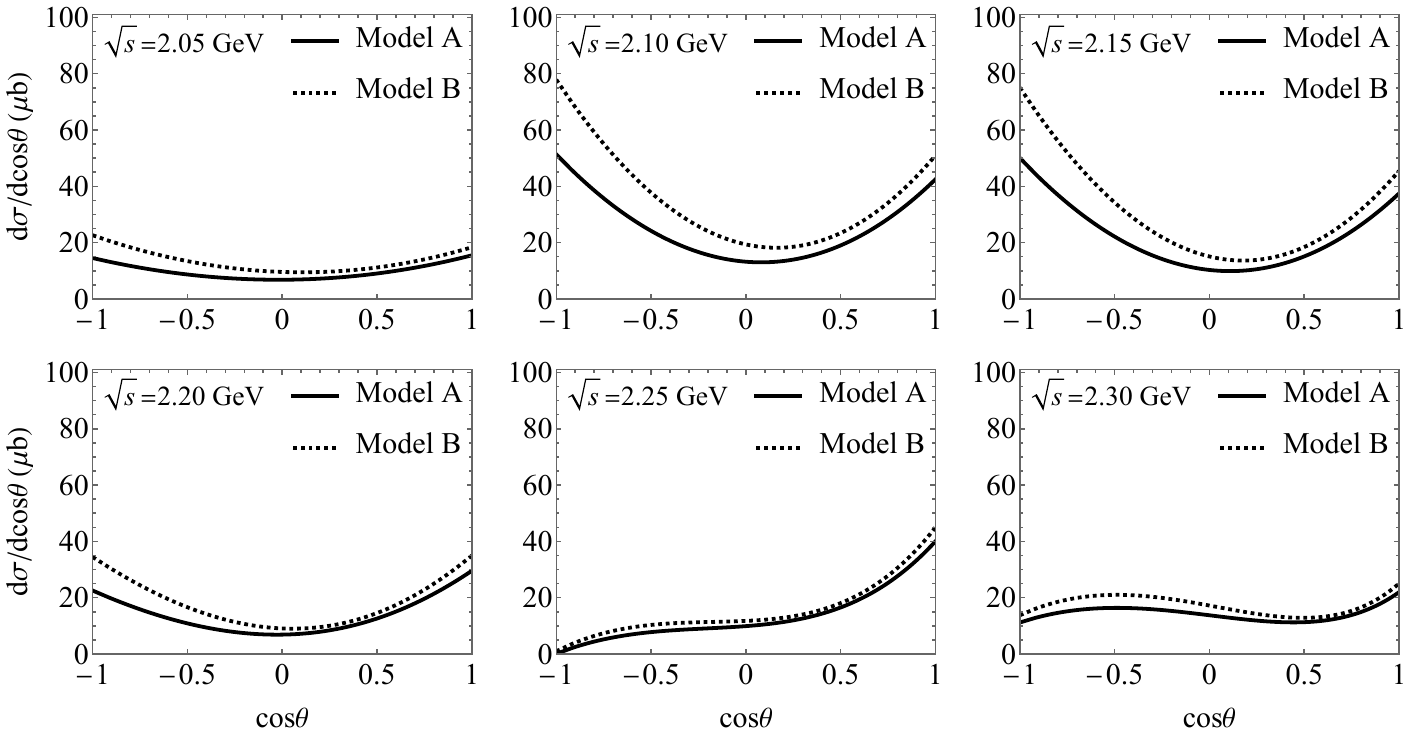}
     \caption{The differential cross sections for $K^{-} p \rightarrow K^{+} \Xi(1530)^{-}$ depending on $\cos \theta$ with several typical values of $\sqrt{s}$. The black solid curves are obtained with the fitted parameters in Table~\ref{Tab.2}, which corresponds to model A. The black dotted curves are obtained with the fitted parameters in Table~\ref{Tab.3}, which corresponds to model B.}
\label{Fig.9}
\end{figure*}

\begin{figure*}[htb]
     \includegraphics[width=175mm]{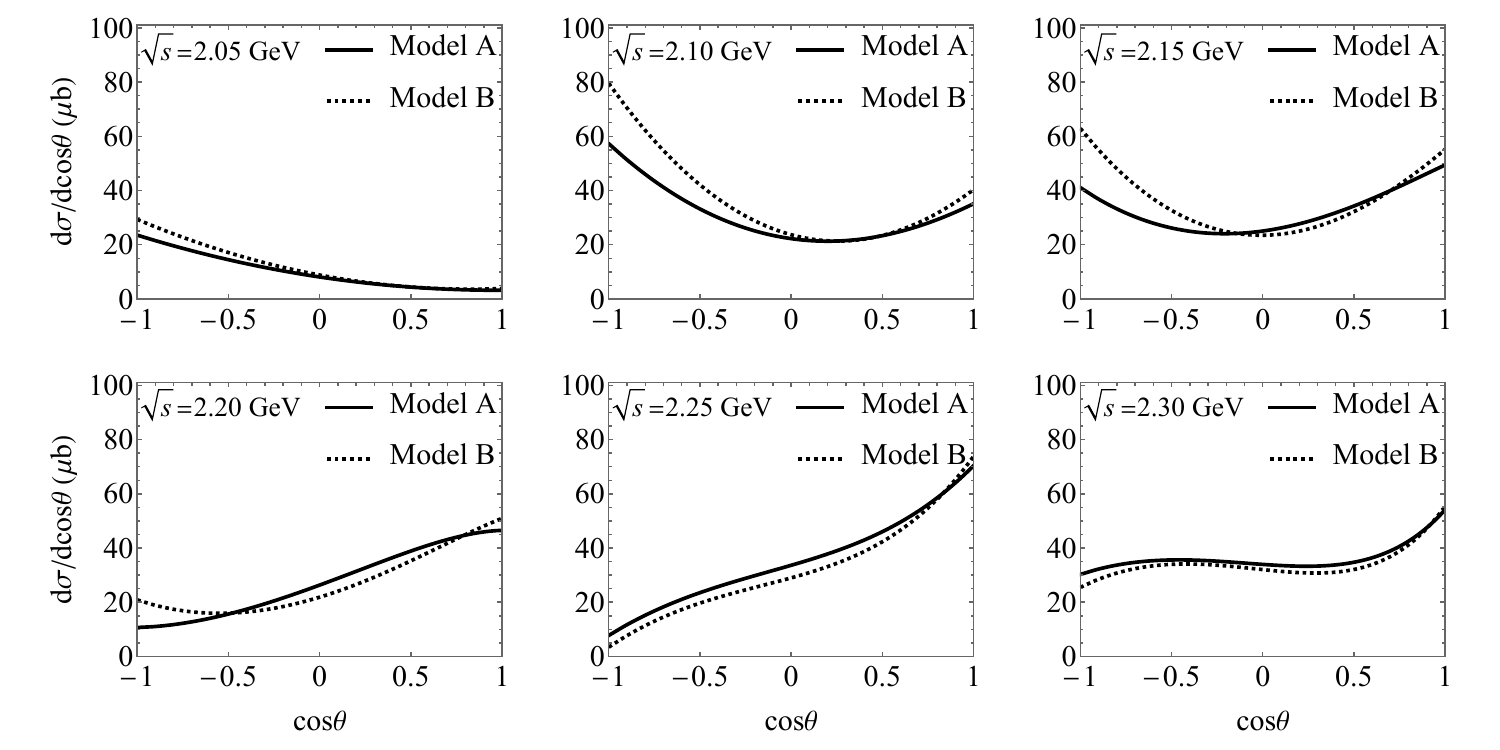}
     \caption{The same as Fig.~\ref{Fig.9} but for the process $K^{-} p \rightarrow K^{0} \Xi(1530)^{0}$.}
\label{Fig.10}
\end{figure*}

In Figs.~\ref{Fig.7} and \ref{Fig.8}, we present the individual contributions from different intermediate states for $K^- p\to K^+ \Xi(1530)^-$ and $K^- p \to K^0 \Xi(1530)^0$, respectively. From the individual contributions, one can find that the resonances $\Sigma(2080)$, $\Sigma(1670)$ and $\Lambda(2110)$ significantly contribute  to  the peak structure around 2.1 GeV. In order to enhance the cross sections for $K^- p \to K^+ \Xi(1530)^-$ in the vicinity of 2.1 GeV, the contributions from these resonances should be increased. By comparing the fitted parameters in Tables~\ref{Tab.2} and \ref{Tab.3}, one can find that in model B the absolute value of the parameters $g_{\Sigma(1670)}$ and $g_{\Lambda(2110)}$ increase by  $33.6\%$ and $20.0\%$, respectively, while the absolute value of $g_{\Sigma(2080)}$ decreases by  $6.8\%$. However, as we analyzed in model A, the contributions for $\Sigma$ resonances are the same for $K^- p\to K^+ \Xi(1530)^-$ and $K^- p \to K^0 \Xi(1530)^0$. In addition, the contributions from above-threshold $\Lambda$ resonance are $s$-channel dominant, indicating the above-threshold $\Lambda$ resonances have similar contributions to $K^- p\to K^+ \Xi(1530)^-$ and $K^- p \to K^0 \Xi(1530)^0$ processes. Thus, the overall enhancement of the cross sections for $K^- p\to K^+ \Xi(1530)^-$ in the vicinity of 2.1 GeV inevitably also enhances the cross sections for $K^- p \to K^0 \Xi(1530)^0$, which lead to a worse global agreement with the experimental data of the cross sections for $K^- p \to K^0 \Xi(1530)^0$ near 2.1 GeV.

 \subsection{Predictions of the differential cross sections for $K^- p \to K \Xi(1530)$}
In addition to the cross sections, the differential cross sections for the involved processes depending on $\cos \theta$ are predicted with the fitted parameters in the present work, where $\theta$ is the angle between the outgoing $\Xi(1530)$ and the ingoing kaon beam direction. The differential cross sections with several typical center-of-mass energies for $K^- p \to K^+ \Xi(1530)^-$ and $K^- p \to K^0 \Xi(1530)^0$ are presented in Figs.~\ref{Fig.9} and ~\ref{Fig.10}, respectively. Here, we take the center-of-mass energy $\sqrt{s}$ as 2.05, 2.10, 2.15, 2.20, 2.25, and 2.30 GeV, respectively. From the figures one can notice that the differential cross sections reach the maximum at the backward angle limit when $\sqrt{s}=2.10$ and 2.15 GeV, while when $\sqrt{s}=2.25$ GeV, the differential cross sections reach the maximum at the forward angle limit. In addition, by comparing model A and model B, one can find that the differential cross sections of model B are generally greater than that of model A, which are consistent with the relevant cross sections. Moreover, the differential cross sections for $K^{-} p \rightarrow K^{0} \Xi(1530)^{0}$ are presented in Fig.~\ref{Fig.10}. Similarly, the black solid curves and black dotted curves correspond to model A and model B, respectively. In most cases, the differential cross sections are similar for $K^- p\to K^+ \Xi(1530)^-$ and $K^- p\to K^0 \Xi(1530)^0$, since the contributions from the $\Sigma$ hyperon are the same for both processes, and those from the above-threshold $\Lambda$ hyperon are also similar. These differential cross sections predicted in the present work could be crucial tests for the present estimations, which should be accessible by the kaon experiments at J-PARC.

\subsection{Cross sections for $K^- p \to K \Xi \pi$ at $p_K=2.87$ GeV}
Since the $\Xi(1530)^{-}$ and $\Xi(1530)^{0}$ states can further decay into $\Xi^{-} \pi^{0}$ and $\Xi^{-} \pi^{+}$, respectively,  one can estimate the cross sections for the cascade processes $K^- p \to K^{+} \Xi(1530)^- \to  K^{+}  \Xi^{-} \pi^{0}$ and $K^- p \to K^{0}\Xi(1530)^0 K^{0} \to \Xi^{-} \pi^{+}$. Experimentally, the authors of  Ref.~\cite{Briefel:1977bp} measured the cross sections for $K^- p \to K^{+} \Xi^{-} \pi^{0}$ and $K^- p \to K^{0} \Xi^{-} \pi^{+}$ to be (5.5$\pm$1.4 ) $\mu$b and $(14.2 \pm 1.8 )\ \mathrm{\mu b}$ at $P_{K}=2.87$ GeV (equivalent to $\sqrt{s}=2.57$ GeV), respectively, where $\Xi^- \pi^0$ and $\Xi^- \pi^+$ are the daughter particles of $\Xi(1530)^-$ and $\Xi(1530)^0$, respectively. Considering the isospin symmetry, the branching ratios of $\Xi(1530)^{-} \to \Xi^{-} \pi^{0}$ and $\Xi(1530)^{0} \to \Xi^{-} \pi^{+}$ are about 33.3$\%$ and 66.6$\%$, respectively~\cite{Flaminio:1979iz}. Then, by combining the cross sections for the two-body processes that have already been obtained in Sec.\ref{sec:MA}, one can predict the cross sections for the three-body processes. For model A, our estimations indicate that the cross sections for $K^{-} p \rightarrow K^{+} \Xi(1530)^{-}$ and $K^{-} p \rightarrow K^{0} \Xi(1530)^{0}$ are about 6.40 and 20.98 $\mu$b at $\sqrt{s}=2.57$ GeV, respectively. Therefore, the cross sections for the charged and neutral three-body processes can be approximately estimated to be 2 and 14 $\mu$b at $\sqrt{s}=2.57$ GeV, respectively. For model B, our estimations indicate that the cross sections for $K^{-} p \rightarrow K^{+} \Xi(1530)^{-}$ and $K^{-} p \rightarrow K^{0} \Xi(1530)^{0}$ are about 10.14 and 21.52 $\mu$b at $\sqrt{s}=2.57$ GeV, respectively. Consequently, the cross sections for the charged and neutral three-body processes are about 3 and 14 $\mu$b at $\sqrt{s}=2.57$ GeV, respectively. To sum up, the estimated results in the present fit are basically consistent with the experimental measurements for the three-body process for both model.

\section{SUMMARY}
The ground $\Xi$ hyperon state, $\Xi(1314)$ has been comprehensively studied over the past years, and the researches on its productions in the $K^-p$ scattering processes have been performed in the literature. However, the research on the productions of $\Xi(1530)$ is limited due to the scarcity of the experimental data. In the present work, we investigate the production of $\Xi(1530)^{0}$ and $\Xi(1530)^-$ in the $K^- p$ scattering process simultaneously, where nine $\Lambda$ and $\Sigma$ hyperons and their resonances are involved in the $s$- and $u$-channel processes. Checking the experimental data, we find that the measured cross sections for $K^- p \to K^+ \Xi(1530)^-$ are inconsistent with each other in the range $\sqrt{s}=[2.087, 2.168]\ \mathrm{GeV}$. Thus, in the present work, we take two different fitting strategies. In model A, the weights of the experimental data are the same, while in model B, the experimental data have different weights. Based on a comparative analysis of the fit results in two models, we conclude that model A demonstrates a better global agreement with the experimental data than model B.

In addition to the cross sections, the individual contributions from different intermediate states to the cross sections for  $K^- p \to K^{+} \Xi(1530)^{-}$ and $K^- p \to K^{0} \Xi(1530)^{0}$ are also estimated with the fitted parameters. Our results indicate that the cross section resulted from the $\Sigma(1193)$ intermediate process is predominant. Moreover, the different cross sections are predicted for both the charged and neutral processes with several typical center-of-mass energies, which could be tested by further experimental measurements at J-PARC in the future.

\section*{ACKNOWLEDGMENTS}
This work is partly supported by the National Natural Science Foundation of China under Grants No. 12175037 and No. 12335001, as well as supported, in part, by the National Key Research and Development Program under Contract No. 2024YFA1610503. PW acknowledges support from the Natural Science Foundation of Jiangsu Province (Grant No. BK20210201), Fundamental Research Funds for the Central Universities, Excellent Scholar Project of Southeast University (Class A), and the Big Data Computing Center of Southeast University.


\end{document}